\documentstyle[aps,prl,epsfig]{revtex}

\begin{document}

\draft

\title{Stationary entanglement between macroscopic mechanical oscillators}

\author{Stefano Mancini$^{1}$, David Vitali$^{1}$, Vittorio Giovannetti$^{2}$,
 and Paolo Tombesi$^{1}$}

\address{$^{1}$INFM, Dipartimento di Fisica, Universit\`a di Camerino,
I-62032 Camerino, Italy  \\
$^{2}$Research Laboratory of Electronics,
MIT - Cambridge, MA 02139, USA
}

\date{\today}

\maketitle

\begin{abstract}
We show that the optomechanical coupling between an optical cavity mode
and the two movable cavity mirrors is able to entangle two 
different macroscopic oscillation modes of the mirrors. This 
continuous variable entanglement is maintained by the light bouncing 
between the mirrors and is robust against thermal noise. In fact, it 
could be experimentally demonstrated using present technology.
\end{abstract}


\section{Introduction}

Entanglement is the most characteristic trait of quantum 
mechanics \cite{SCH}. An entangled state of a system consisting of 
two subsystems cannot be described as a product (or a statistical 
mixtures of products) of the quantum states of the two subsystems. In 
such a state, the system is inseparable and each component does not 
have properties independent of the other components. The nonlocal character of 
entangled states is at the basis of many paradoxes \cite{EPR}, and of 
the deep difference between the quantum and the classical world. The 
fundamental role of entanglement has been reemphasized in recent 
years after the discovery that it represents an unvaluable resource
for quantum information processing \cite{ike}. In fact, entanglement is at the 
basis of secure quantum key distribution schemes \cite{ekert}, of 
quantum teleportation \cite{telep}, and of the speed-up provided by 
some quantum algorithms \cite{ekejos}. It is generally believed that 
entanglement can be found only in situations involving a small number of 
microscopic particles. For example, a given amount of entanglement is 
present between two different spins in the thermal equilibrium 
state of a system of many interacting spins (the so-called thermal or 
natural entanglement \cite{zana}). However, for quantum information 
processing, it is the deterministic generation and manipulation
of entanglement which is of paramount importance, and in these last years a 
number of impressive experiments has demonstrated the controlled 
generation of entangled states of two \cite{2-ent}, three \cite{3-ent} 
and four \cite{4-ent} particles. Moreover, since entanglement is one of the 
distinguishing features of the quantum world, it is also fundamental 
to understand how far it can be extended into the macroscopic domain. 
This is important not only to better establish how the macroscopic
classical world emerges from the microscopic one ruled by quantum mechanics 
\cite{zur}, but also for application purposes. For example, entangled 
spin-squeezed states of atomic samples
are known to improve the precision of frequency 
measurements \cite{wine}, and the accuracy improves with increasing 
number of entangled atoms. A related question is to establish if and 
how two macroscopic degrees of freedom of two different objects can 
be entangled. With this respect, a striking achievement has been 
recently shown in \cite{juls}, where the entanglement between the 
spin states of two separated Cs gas samples containing about 
$10^{12}$ atoms has been demonstrated. At the same time we  
proposed a feasible experiment \cite{prlno} in which even a more 
macroscopic entanglement between the oscillating modes of two mirrors 
with an effective mass of some milligrams can be generated by the 
radiation pressure of the light bouncing between them (see also
\cite{zheng} for a different and extremely idealized model 
for the preparation of motion entangled states of two cavity mirrors). 
The continuous variable entanglement between two mechanical modes could be used 
to improve the detection of weak classical forces in optomechanical 
devices as atomic force microscopes or gravitational wave detectors 
\cite{stefa,milb1}.

In this paper we analyze in more detail and further develop the 
proposal of \cite{prlno}. In fact, Ref.~\cite{prlno} restricted to 
the case of identical cavity mirrors, i.e., considered, for each 
mirror, a single oscillation mode with identical effective mass,
optomechanical coupling, damping rate and, above all, identical 
resonance frequency. However, \cite{prlno} showed that the entanglement 
is present only within a small bandwidth around the mechanical 
resonance, and since in practice two mirrors are never exactly 
identical, it is important to establish the conditions under which 
entanglement can be generated between two mechanical modes with 
different resonance frequencies, and its dependence on the frequency 
mismatch.

In Section II we describe the optomechanical system under study in 
terms of quantum Langevin equations. In Section III we solve the 
dynamics of the system in the frequency domain,
and then we characterize in detail the entanglement between the two mirrors. 
Section IV is for concluding remarks.

\section{The system}

We consider an optical ring cavity in which two perfectly reflecting   
mirrors can both oscillate under the 
effect of the radiation pressure force (see Fig.~1).
The motion of each mirror is the result of the 
excitation of many oscillation modes, either external \cite{DOR,TITTO}
or internal \cite{HADJAR,raab}. The former is important for suspended 
mirrors since the excitation of pendulum modes of the suspension 
system leads to global displacements of the mirror. The latter 
corresponds to deformations of the mirror surface due to the 
excitation of internal acoustic modes of the substrate.
These various degrees of freedom have however different resonance 
frequencies and one can select the mechanical response of a single 
particular mode by using a bandpass filter
in the detection circuit \cite{hadjar2}. For this reason we shall 
consider a single mechanical mode for each mirror, which will be 
therefore described as a simple harmonic oscillator. Since 
we shall consider two mirrors with similar design, the two modes will 
be characterized by different, but 
quite close, values for the frequencies, $\Omega_{1}$ and 
$\Omega_{2}$, and for the effective masses, $m_{1}$ and $m_{2}$. 

The optomechanical coupling between the mirrors and
the cavity field is realized by the radiation pressure. The 
electromagnetic field exerts a force on a movable mirror which is 
proportional to the intensity of the field, which, at the same time, 
is phase-shifted by a quantity proportional to the
the mirror displacement from the equilibrium position.
In the adiabatic limit in which the mirror frequency is much smaller 
than the cavity free spectral range $c/(2\sqrt{2}L)$ ($L$
is the diagonal of the square optical path in the cavity, see Fig.~1), 
one can focus on one cavity mode only (with annihilation 
operator $b$ and frequency $\omega_{b}$), because
photon scattering into other modes can be neglected \cite{LAW}. 
One gets the 
following Hamiltonian \cite{HAMI}
\begin{eqnarray}
\label{H}
{\cal H}&=& \hbar\omega_{b} b^{\dag}b
+\sum_{i=1}^{2}\frac{\hbar\Omega_{i}}{2}\left(
p_{i}^{2}+q_{i}^{2}\right) \\
&&
-\hbar b^{\dag} b \sum_{j=1}^{2}(-1)^{j}G_{j}q_{j}
+i\hbar\sqrt{\gamma_b}\left(\beta^{in}e^{-i\omega_{b0}t}b^{\dag}
-\beta^{in\,*}e^{i\omega_{b0}t}b\right) \nonumber \,,
\end{eqnarray}
where $q_{i}$ and $p_{i}$ are the dimensionless 
position and momentum operators of the mirrors with 
$\left[q_{i},p_{j}\right]=i\delta_{ij}$, 
$G_{j}=(\omega_{b}/2L)\sqrt{\hbar/m_{j}\Omega_{j}}$ $(j=1,2)$ are the 
optomechanical coupling constants, and the last terms in Eq.~\ref{H} 
describe the laser driving the cavity mode, characterized by a 
frequency $\omega_{b0}$ and a power 
$P_b^{in}= \hbar \omega_{b0} |\beta^{in}|^{2}$
($\gamma_b$ is the cavity mode linewidth).

A detailed analysis of the problem, however, must include 
photon losses, and the
thermal noise on the mirrors.
It means that the interaction of the optical mode with its 
reservoir and the effect of thermal fluctuations on the two mirrors, not 
considered in Hamiltonian (\ref{H}), must be added. This 
can be accomplished in the 
standard way \cite{milwal,VIT}. 
We neglect instead all the technical sources of noise, i.e., 
we shall assume that the driving laser is stabilized
in intensity and frequency, also because recent experiments have shown
that classical laser noise can be made negligible in the relevant 
frequency range \cite{TITTO,HADJAR}.
The full quantum dynamics of the system can be exactly described 
by the following nonlinear 
Langevin equations (in the interaction picture with respect to 
$\hbar \omega_{b}b^{\dagger} b$) 
\begin{equation}\label{NONLINEQS}
\begin{array}{l}
\dot{b} = i(\omega_{b0}-\omega_b) b - i b (G_{1} q_1-G_{2}q_2) -  
\frac{\gamma_b}{2} b+\sqrt{\gamma_b} \left(b^{in}+\beta_{in}\right)\,,
\\ 
\dot{q}_j=\Omega_{j} p_j \,,
\\ 
\dot{p}_j=-\Omega_{j} q_j
+(-)^j G_{j} b^{\dagger} b -\Gamma_{j} p_j+\xi_j \,,
\end{array} 
\end{equation}
where $\Gamma_{j}$ ($j=1,2$) are the mechanical damping rates of the 
mechanical modes, $b^{in}(t)$ 
represent the vacuum white noise operator 
at the cavity input \cite{milwal}, and 
the Langevin noise operators for the
quantum Brownian motion 
of the mirrors are $\xi_j(t)$.
The non-vanishing noise correlations are
\begin{eqnarray}\label{NOISE}
&&\langle b^{in}(t) b^{in\,\dag}(t') 
\rangle = 
\delta(t-t')\,,
\\
&&\langle {\xi_j}(t) {\xi_k}(t') \rangle =
\delta_{j,k}\int_{0}^{\infty} \, d\omega \,  
\frac{\Gamma_{j}\omega}{2\Omega_{j}}
\left[\coth\left(
\frac{\hbar\omega}{2k_BT}\right)\cos\left[\omega(t-t')\right]-
i\sin\left[\omega(t-t')\right] \right] 
\,,\nonumber
\end{eqnarray}
where $k_B$ is the Boltzmann constant and $T$ the 
equilibrium temperature (the two mirrors are considered
in equilibrium with their respective bath at the same 
temperature). Notice that the used approach for the Brownian
motion is quantum mechanical consistent at every temperature
\cite{VIT}.

We consider the situation when the driving field is
very intense. Under this condition, the system is characterized by a
semiclassical steady state with 
the internal cavity mode in a coherent
state $|\beta\rangle $, and a displaced equilibrium position for the 
mirrors. The steady state values are obtained by taking the 
expectation values of 
Eqs.~(\ref{NONLINEQS}), factorizing them and setting 
all the time derivatives to zero. One gets
\begin{eqnarray}
\begin{array}{l}
\langle q_j \rangle_{ss}
= (-)^{j}G_{j}|\beta|^2/\Omega_{j},
\\ 
\langle p_j \rangle_{ss} = 0,
\\ 
\beta \equiv \langle b \rangle_{ss}
=\sqrt{\gamma_b} \beta^{in}/
\left(\gamma_b/2-i\Delta_b\right),
\end{array}
\end{eqnarray}
where $\Delta_b\equiv\omega_{b0}-\omega_b
-G_{1}\langle q_1 \rangle_{ss}+G_{2}\langle q_2 \rangle_{ss}$,
is the cavity mode detuning.

Under these semiclassical conditions, the dynamics is well described 
by linearizing Eqs.~(\ref{NONLINEQS}) around the steady state. If we now
use the same symbols for the operators describing the quantum 
fluctuations around the steady state, we get
the following linearized quantum Langevin equations
\begin{equation}\label{LINEQS}
\begin{array}{l} 
\dot{b} = i\Delta_b b-i \beta (G_{1}q_1-G_{2}q_2) -  
\frac{\gamma_b}{2} b+\sqrt{\gamma_b} b^{in}\,,
\\ 
\dot{q}_j=\Omega_{j} p_j \,,
\\ 
\dot{p}_j=-\Omega_{j} q_j
+(-)^j G_{j} (\beta^* b+\beta b^{\dag})-\Gamma_{j} p_j+\xi_j \,.
\end{array} 
\end{equation}

\section{Entanglement characterization}

The time evolution of the system can be easily obtained by solving the 
linear quantum Langevin equations (\ref{LINEQS}). However, as it 
happens in quantum optics for squeezing (see for example 
\cite{milwal}), it is more convenient to study the system dynamics in 
the frequency domain. In fact, it is possible that, due to the effect 
of damping, and thermal and quantum noise, the two mechanical modes of 
the mirrors are never entangled in time, i.e., there is no time 
instant in which the reduced state of the two mechanical modes is 
entangled, unless appropriate (but difficult to prepare) initial 
conditions of the whole system are considered. Entanglement can be 
instead always present at a given frequency. In fact, the two mirrors 
constitute, for each frequency, a continuous variable bipartite system 
which, in a given frequency bandwidth, can be in an entangled state.
The Fourier analysis refers to the quantum fluctuations around the 
semiclassical steady state discussed in the preceding Section, and 
the eventual entanglement found at a given frequency would refer to a 
stationary state of the corresponding spectral modes, 
maintained by the radiation mode, 
and which decays only when the radiation is 
turned off. The spectral analysis is more convenient also because
in such systems the dynamics is experimentally 
better studied in frequency rather than in time. The same kind of 
spectral analysis of the nonlocal properties of a bipartite 
continuous variable system has been already applied in Ref.~\cite{kimb}
which demonstrated the EPR nonlocality between two optical beams of a 
nondegenerate parametric amplifier, following the suggestion of 
\cite{REID}.

Performing the Fourier transform of Eqs.~(\ref{LINEQS}), 
one easily gets for the mechanical modes operators ($j=1,2$)
\begin{eqnarray}
&& q_{j}(\omega)={\cal B}_{j}(\omega)b_{in}(\omega)
+{\cal B}_{j}^{*}(-\omega)b_{in}^{\dag}(-\omega)
+\Xi_{j,1}(\omega) \xi_{1}(\omega)+\Xi_{j,2}(\omega) \xi_{2}(\omega) 
\label{qspe} \\
&&p_{j}(\omega)=-i\frac{\omega}{\Omega_{j}} q_{j}(\omega)\,,
\label{pspe}
\end{eqnarray}
where
\begin{mathletters}
\label{techni}
\begin{eqnarray}
{\cal B}_{j}(\omega)&=&
(-)^{j}\frac{1}{{\cal D}(\omega)}\left[
\frac{1}{\Omega_{3-j}\chi_{3-j}(\omega)}\right]
\left[\frac{\sqrt{\gamma_{b}}G_{j}\beta^{*}}{\frac{\gamma_{b}}{2}
-i\left(\Delta_{b}+\omega\right)}\right]\,,
\\
\Xi_{j,k}(\omega)&=&\frac{1}{{\cal D}(\omega)}
\left\{
\frac{1}{\Omega_{3-j}\chi_{3-j}(\omega)}\delta_{j,k}
\right.
\nonumber\\
&&\left.
-i G_{3-j}G_{3-k}|\beta|^{2}\left[
\frac{1}{\frac{\gamma_{b}}{2}-i\left(\Delta_{b}+\omega\right)}
-\frac{1}{\frac{\gamma_{b}}{2}+i\left(\Delta_{b}-\omega\right)}
\right]\right\}\,,
\\
{\cal D}(\omega)&=&
\frac{1}{\Omega_{1}\Omega_{2}\chi_{1}(\omega)\chi_{2}(\omega)}
\nonumber\\
&&-i|\beta|^{2}\left[\frac{G_{1}^{2}}{\Omega_{2}\chi_{2}(\omega)}
+\frac{G_{2}^{2}}{\Omega_{1}\chi_{1}(\omega)}\right]
\left[
\frac{1}{\frac{\gamma_{b}}{2}-i\left(\Delta_{b}+\omega\right)}
-\frac{1}{\frac{\gamma_{b}}{2}+i\left(\Delta_{b}-\omega\right)}
\right]\,,
\end{eqnarray}
\end{mathletters}
and $\chi_{j}(\omega)=[\Omega_{j}^{2}-\omega^{2}
-i\omega\Gamma_{j}]^{-1}$ is the mechanical susceptibility
of mode $j$.
Notice that $\Xi_{j,k}^{*}(\omega)=\Xi_{j,k}(-\omega)$ and
${\cal D}^{*}(\omega)={\cal D}(-\omega)$, but
${\cal B}^{*}(\omega)\ne{\cal B}(-\omega)$.

The simplest way to establish the parameter region where 
the oscillation modes of the two cavity mirrors are entangled is to 
use one of the sufficient criteria for entanglement of continuous 
variable systems already existing in the literature. These criteria are 
inequalities which have to be satisfied by the product 
\cite{prlno,tan,reid2} or the sum \cite{DUAN,SIMON} of variances of 
appropriate linear combinations of the rescaled position and momentum 
operators of the continuous variable systems. These criteria are 
usually formulated in terms of Heisenberg operators at the same time 
instant, satisfying the usual commutation relations 
$\left[q_{j}(t),p_{k}(t)\right]=i\delta_{jk}$ \cite{tan,DUAN,SIMON}, but 
they can be adapted to their Fourier transform, as long as the 
commutators between the frequency-dependent continuous variable 
operators are still a c-number \cite{prlno}. This condition is 
satisfied in the present case thanks to the linearity of the Fourier 
transform and to the linear dynamics of the fluctuations (see 
Eqs.~(\ref{LINEQS})), implying that the commutators are always 
c-number frequency-dependent functions. 

The paradigmatic entangled state for continuous variable systems is 
the state considered by Einstein, Podolski and Rosen in their famous 
paper \cite{EPR}, i.e., the simultaneous eigenstate of the 
relative distance $q_{1}-q_{2}$ and of the total momentum 
$p_{1}+p_{2}$. In an entangled state of this kind,
the variances of these two 
operators are both small and it is therefore natural to use them. 
Defining $u=q_{1}-q_{2}$ and 
$v=p_{1}+p_{2}$, an inseparability criterion
for the sum of the variances in the 
case of arbitrary c-number commutators is \cite{DUAN}
\begin{equation}  \label{e3}
\left\langle \left( \Delta u\right) ^{2}\right\rangle 
+\left\langle \left( \Delta v\right) ^{2}\right\rangle 
< 2 |\langle[q_1,p_1]\rangle|^2 \,,
\end{equation}
while that for the product of variances is \cite{prlno,tan,reid2}
\begin{equation}  \label{e4}
\left\langle \left( \Delta u\right) ^{2}\right\rangle 
\left\langle \left( \Delta v\right) ^{2}\right\rangle 
< |\langle[q_1,p_1]\rangle|^2 \,.
\end{equation}
It is easy to see that the condition (\ref{e3})
implies condition (\ref{e4}), which means that the product 
criterion (\ref{e4}) is easier to satisfy, and for this reason we 
shall consider only the latter from now on. 
Furthermore, the product criterion (\ref{e4}) allows us to establish 
a connection with Refs.~\cite{REID}, 
which showed that when the inequality
\begin{equation}\label{INFER}
\left\langle \left( \Delta u\right) ^{2}\right\rangle 
\left\langle \left( \Delta v\right) ^{2}\right\rangle 
< \frac{1}{4}|\langle[q_1,p_1]\rangle|^2\,,
\end{equation}
is satisfied, an EPR-like paradox arises \cite{EPR,epl}, based on the 
inconsistency between quantum mechanics and local realism, which has 
been then experimentally confirmed in \cite{kimb}.
The sufficient condition for inseparability of 
Eq.~(\ref{e4}) is weaker than condition (\ref{INFER}), but this is not 
surprising, since entangled states are only a necessary condition for the 
realization of an EPR-like paradox (see however the recent
paper \cite{reid2} where it is shown
that the weaker inseparability sufficient condition (\ref{e4})
can be considered as a marker of the existence of generalized,
weaker, EPR correlations).

In order to apply the inseparability criterion (\ref{e4})
in the frequency domain, we have to make the 
frequency dependent operators $q_{j}(\omega)$ and $p_{j}(\omega)$ 
Hermitian, i.e., to consider the Hermitian component
\begin{equation}
{\cal R}_{{\cal O}}(\omega)=\frac{{\cal O}(\omega)+{\cal 
O}(-\omega)}{2}
\end{equation}
for any operator ${\cal O}(\omega)$. Using the fact 
that $\langle q_{j}(\omega)\rangle = \langle p_{j}(\omega)\rangle = 
0$, $j=1,2$ and $\forall \omega$ because they are associated to 
fluctuations around the semiclassical steady state, Eq.~(\ref{e4}) 
therefore becomes
\begin{equation}
\langle {\cal R}_{q_{1}-q_{2}}^{2} \rangle
\langle {\cal R}_{p_{1}+p_{2}}^{2} \rangle <
\left|\langle \left[
{\cal R}_{q_{1}},{\cal R}_{p_{1}}\right]\rangle\right|^{2}\,,
\end{equation}
which suggests the following definition of
degree of entanglement for the mechanical 
oscillation modes at frequency $\omega$
of the two cavity mirrors \cite{prlno}
\begin{equation}
\label{degree}
E(\omega)=\frac{\langle {\cal R}_{q_{1}-q_{2}}^{2} \rangle
\langle {\cal R}_{p_{1}+p_{2}} ^{2}\rangle}
{\left|\langle \left[
{\cal R}_{q_{1}},{\cal R}_{p_{1}}\right]\rangle\right|^{2}}\,,
\end{equation}
which is a marker of entanglement whenever $E(\omega) < 1$.

Using Eqs.~(\ref{techni}) it is possible to derive the analytic 
expression of $E(\omega)$, which is however very cumbersome.
The two variances in the numerator of (\ref{degree}) are
\begin{eqnarray}
\langle {\cal R}_{q_{1}-q_{2}}^{2} \rangle &=&
\frac{1}{4}\left\{
|{\cal B}_{1}(\omega)-{\cal B}_{2}(\omega)|^{2} 
+|{\cal B}_{1}(-\omega)-{\cal B}_{2}(-\omega)|^{2} \right.
\nonumber\\
&&\left.
+{\cal N}_{1}(\omega) |\Xi_{1,1}(\omega)-\Xi_{2,1}(\omega)|^{2}
+{\cal N}_{2}(\omega) |\Xi_{1,2}(\omega)-\Xi_{2,2}(\omega)|^{2}
\right\}\,,
\end{eqnarray}
\begin{eqnarray}
\langle {\cal R}_{p_{1}+p_{2}}^{2} \rangle &=&
\frac{1}{4}\left(\frac{\omega}{\Omega_{1}}\right)^{2}
\left\{
|{\cal B}_{1}(\omega)|^{2}+|{\cal B}_{1}(-\omega)|^{2} 
+{\cal N}_{1}(\omega) \left[|\Xi_{1,1}(\omega)|^{2}
+|\Xi_{2,1}(\omega)|^{2}\right]\right\}
\nonumber\\
&&+\frac{1}{4}\left(\frac{\omega}{\Omega_{2}}\right)^{2}
\left\{
|{\cal B}_{2}(\omega)|^{2}+|{\cal B}_{2}(-\omega)|^{2} 
+{\cal N}_{2}(\omega) \left[|\Xi_{1,2}(\omega)|^{2}
+|\Xi_{2,2}(\omega)|^{2}\right]\right\}
\nonumber\\
&&+\frac{1}{4}\left(\frac{\omega^{2}}{\Omega_{1}\Omega_{2}}\right)
\left\{
{\cal B}_{1}(\omega){\cal B}_{2}^{*}(\omega)
+{\cal B}_{1}(-\omega){\cal B}_{2}^{*}(-\omega)
+{\cal B}_{1}^{*}(\omega){\cal B}_{2}(\omega)
+{\cal B}_{1}^{*}(-\omega){\cal B}_{2}(-\omega)
\right.
\nonumber\\
&&\left.
+{\cal N}_{1}(\omega) \left[\Xi_{1,1}(\omega)\Xi_{2,1}(-\omega)
+\Xi_{1,1}(-\omega)\Xi_{2,1}(\omega)
\right]\right.
\nonumber\\
&&\left.
+{\cal N}_{2}(\omega) \left[\Xi_{1,2}(\omega)\Xi_{2,2}(-\omega)
+\Xi_{1,2}(-\omega)\Xi_{2,2}(\omega)
\right]\right\}\,,
\end{eqnarray}
with ${\cal N}_{j}(\omega)=\omega(\Gamma_{j}/\Omega_{j})
\coth(\hbar\omega/2 k_{B}T)$, while the commutator in the denominator
of (\ref{degree}) is given by
\begin{equation}
\langle \left[
{\cal R}_{q_{1}},{\cal R}_{p_{1}}\right]\rangle=
\frac{i}{2}\frac{\omega}{\Omega_{1}}
\left\{|{\cal B}_{1}(\omega)|^{2}-|{\cal B}_{1}(-\omega)|^{2}
-\omega\frac{\Gamma_{1}}{\Omega_{1}}
\left[|\Xi_{1,1}(\omega)|^{2}+|\Xi_{1,2}(\omega)|^{2}\right]
\right\}\,.
\end{equation}
In Figs.~2-4 we have studied the behaviour of $E(\omega)$ as a 
function of frequency and temperature, for different values of the 
difference between the two resonance frequencies of the mechanical 
modes, $\Omega_{1}-\Omega_{2}$. This is an important parameter because 
we have seen in \cite{prlno} that in the case of identical mirrors,
the two mechanical modes are entangled only within a small bandwidth 
around the mechanical resonance. Since in practice the two mirrors 
will never be exactly identical, it is important to establish if the 
macroscopic entanglement is able to tolerate a certain amount
of frequency mismatch. For the 
other parameter values we have considered an experimental situation 
comparable to that of Refs.~\cite{HADJAR,hadjar2,pinard}, where the 
studied mirror oscillation mode is a Gaussian acoustic mode. We have 
therefore considered a cavity driven by a laser working at
$\lambda = 810$ nm and power $P_{b}^{in}=1$ W,
with length $L=1$ mm, detuning $\Delta_{b} =6$ MHz,
optical finesse ${\cal F}=25000$, 
yielding a cavity decay rate $\gamma_{b}=6$ MHz.
The mechanical modes have been taken with effective mass 
$m_{1}=m_{2}=23$ mg, damping rates $\Gamma_{1}=\Gamma_{2}=1$ Hz, and
$\Omega_{1}=1$ MHz, while we have changed the values of $\Omega_{2}$ 
around those of $\Omega_{1}$. These choices yield for the optomechanical 
couplings $G_{1} \simeq G_{2} \simeq 2.5$ Hz.

Fig.~2 shows $E(\omega,T)$ for no frequency mismatch, $\Omega_{1}= 
\Omega_{2}$, Fig.~3 refers to the case with 
$\Omega_{2}-\Omega_{1}=10 $ Hz, and Fig.~4 refers to the case with 
$\Omega_{2}-\Omega_{1}=20 $ Hz. In all cases, the region of the 
$\omega,T$ plane where the two mechanical modes are entangled is 
centered in the middle of the two mechanical resonances, i.e., $E(\omega,T)$
always achieves its minimum at $\omega = (\Omega_{1}+\Omega_{2})/2$. 
The frequency bandwidth of the entanglement region rapidly decreases 
with increasing temperature, so that, with the chosen parameter 
values, entanglement disappears above $T \simeq 4$ K. As expected, 
the $\omega,T$ region where the two mirrors are entangled  
becomes smaller for increasing frequency mismatch (compare the three 
figures). Nonetheless these results are extremely interesting because
they clearly show the possibility to entangle two {\em macroscopic 
oscillators} (with an effective mass of $23$ mg) in a stationary way, 
using present technology.
In fact, the two modes are still clearly entangled at $T = 2$ K and with 
$\Omega_{2}-\Omega_{1}=10 $ Hz (ten times larger than
the width of the mechanical resonance peaks, see Fig.~3), 
while one is forced to go below $T = 2$ K
when the frequency mismatch is equal to $20$ Hz (see Fig.~4).

Differently from temperature and frequency mismatch, and
as it can be seen from the involved analytical expression above, 
it is difficult to determine how the degree of 
entanglement depends upon the other parameters. It can only be verified 
that, as expected, entanglement improves with increasing mechanical 
quality factor $Q_{j}=\Omega_{j}/\Gamma_{j}$ and that it strongly 
improves with increasing the effective optomechanical coupling 
constant, which is given by $\beta G_{j}$ (see Eqs.~(\ref{LINEQS})).
This shows that for achieving even a more macroscopic entanglement 
(i.e., larger masses), one has to use smaller cavities and, above 
all, larger optical power. The fundamental importance of the effective coupling 
constant $\beta G_{j}$ also helps us to show which kind of entangled 
state of the two mirrors is generated by the radiation pressure. In 
fact, when the cavity mode intensity becomes larger and larger, the 
optomechanical interaction tends to project
the two mechanical modes onto an approximate 
eigenstate of $G_{1}q_{1}-G_{2}q_{2}$ (see Eqs.~(\ref{H}) and
(\ref{LINEQS})), 
which, since in our case it is 
$G_{1} \simeq G_{2}$, is essentially equivalent to the
relative distance $q_{1}-q_{2}$.
The two oscillators occupy a state that, like a
standard EPR state, has a very small variance of the relative
distance $u=q_1-q_2$. On the other hand, since the
radiation pressure does not have analogous effects on the total momentum
$v=p_1 + p_2$, the state of the mirrors does not exhibit such a small
value for the variance $\left\langle \left( \Delta v\right) ^{2}\right\rangle$
as the standard EPR state does.
Nonetheless, at large optical intensities,  
as shown by the product criterion of 
Eq.~(\ref{e4}), the effect of the radiation 
pressure force on the relative distance is sufficient 
to entangle the two macroscopic oscillator modes. Moreover, as it can 
be seen from Figs.~2-4, the degree of entanglement $E(\omega)$ lies 
even below $1/4$ at sufficiently low temperatures, allowing therefore in 
principle also an experimental test of EPR nonlocality with 
macroscopic oscillators, on the basis of the inequality (\ref{INFER}) 
of Refs.~\cite{REID}.

\section{Conclusions}

We have shown how the optomechanical coupling realized by the 
radiation pressure of an optical mode of a ring cavity 
is able to entangle two macroscopic collective 
oscillation modes of two cavity mirrors. Using parameter values 
corresponding to already performed experiments involving an optical cavity 
mode coupled to an acoustic mode of the mirror (with an effective mass 
of many milligrams) we have shown that an 
appreciable entanglement is achievable at temperatures of some Kelvin.
This continuous variable entanglement is established at a given 
frequency, between the 
spectrally decomposed oscillation modes of the two mirrors (see also 
Refs.~\cite{kimb,REID} for an analogous spectral analysis of the 
nonlocal properties of the beams of a nondegenerate optical amplifier). 
One has a stationary entanglement, which is maintained by the 
strongly driven cavity mode as long as it is turned on. Using the degree 
of entanglement $E(\omega)$ of Eq.~(\ref{degree}), suggested by the 
inseparability condition of Eq.~(\ref{e4}), we have seen 
that the entanglement is more robust when the two mechanical resonance 
frequencies are equal (Fig.~2), but that it tolerates a resonance 
frequency mismatch of tens of Hz, much larger than the width of the 
resonance peaks. The best entanglement is always achieved in the middle
of the two mechanical resonances and the frequency bandwidth of the 
entanglement parameter region rapidly decreases with decreasing 
optomechanical coupling and increasing temperatures.

This continuous variable entanglement between two macroscopic 
collective degrees of freedom can be experimentally measured using 
for example the three-cavity scheme described in detail in 
\cite{prlno}. In such a scheme, a ring cavity is 
supplemented with two other external cavities, each measuring the 
spectral components $q_{j}(\omega)$ and $p_{j}(\omega)$ of each 
mirror oscillation mode via homodyne detection. With these 
measurements, it is possible to obtain both variances 
$\langle {\cal R}_{q_{1}-q_{2}}^{2} \rangle$ and 
$\langle {\cal R}_{p_{1}+p_{2}}^{2} \rangle$. As it has been verified 
in \cite{prlno}, if the driving power of the meter cavities is much 
smaller than the driving power of the ``entangler'' cavity mode, the 
two additional cavities do not significantly modify the entanglement 
dynamics. A simplified detection scheme, involving less than three 
cavities is currently investigated. In fact, the homodyne detection 
of the entangler mode $b$ provides direct information on the relative 
distance between the mirror modes $q_{1}-q_{2}$. The measurement of 
the total momentum quadrature $p_{1}+p_{2}$ could be then achieved 
using the result of this homodyne detection and that of the 
homodyne measurement of the motion of a single mirror provided by a 
second ``meter'' cavity mode. It is however possible that an even 
simpler detection scheme exists, using the entangler cavity mode only.

Another important aspect which has to be taken into account
is that the motion of each mirror is the 
superposition of many oscillation modes with different resonance 
frequencies. We can safely verify the entanglement between the two
considered oscillation modes provided that the other modes of the two 
mirrors are sufficiently far away in frequency so that their 
contribution at the analysed frequencies is negligible.
Moreover, the above analysis also applies, almost unmodified,
to the case when the two modes belong to the same mirror.

The possibility to prepare entangled state of two macroscopic 
degrees of freedom is not only conceptually important for better 
understanding the relation between the macroscopic world ruled by 
classical mechanics and the quantum mechanical microscopic substrate, 
but it may also prove to be useful for some applications.
For example, Ref.~\cite{stefa} has showed that entangled states of the 
kind studied here could improve the detection of weak mechanical 
forces acting on the mirrors as those due to gravitational waves 
\cite {GRAV}.

\begin{figure}[t]
\centerline{\epsfig{figure=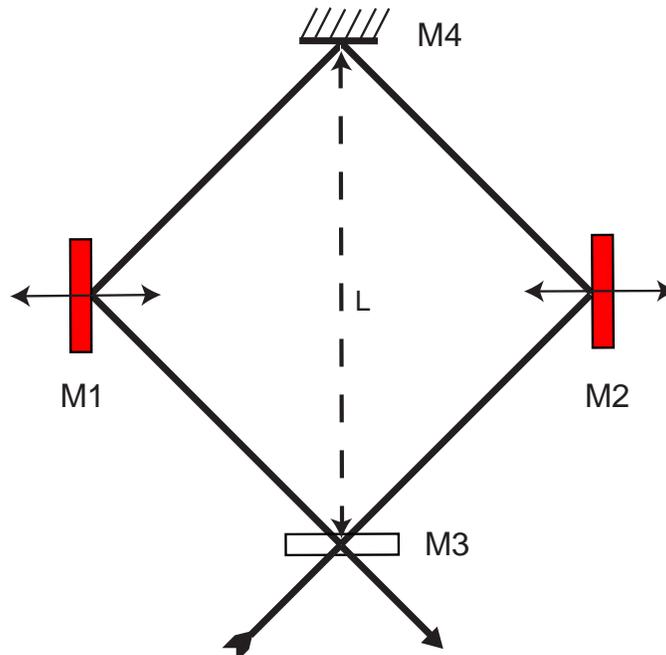,width=3.5in}}
\caption{ 
Schematic description of the system under study.
$L$, being the equilibrium distance between the movable mirrors 
$M1$, $M2$,
is assumed to also be the distance between the fixed mirrors
$M3$, $M4$.
The mirror $M3$ represents the input-output port of the cavity. 
}
\label{fig1}
\end{figure}

\begin{figure}[t]
\centerline{\epsfig{figure=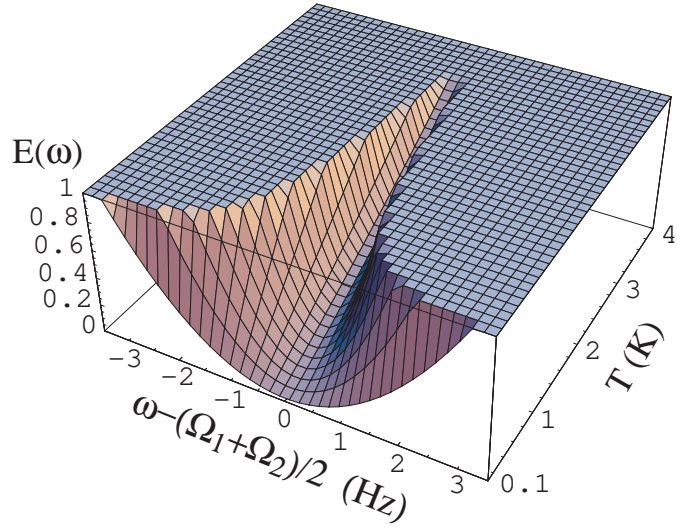,width=3.5in}}
\caption{ 
Degree of entanglement $E(\omega)$ 
of Eq.~(\protect\ref{degree}) versus 
frequency and temperature $T$, in the case of equal
mechanical resonance frequencies, $\Omega_{1}=\Omega_{2}=1$ MHz.
The plot has been cut at $E(\omega)=1$.
The other parameter values are in the text.
}
\label{fig2}
\end{figure}

\begin{figure}[t]
\centerline{\epsfig{figure=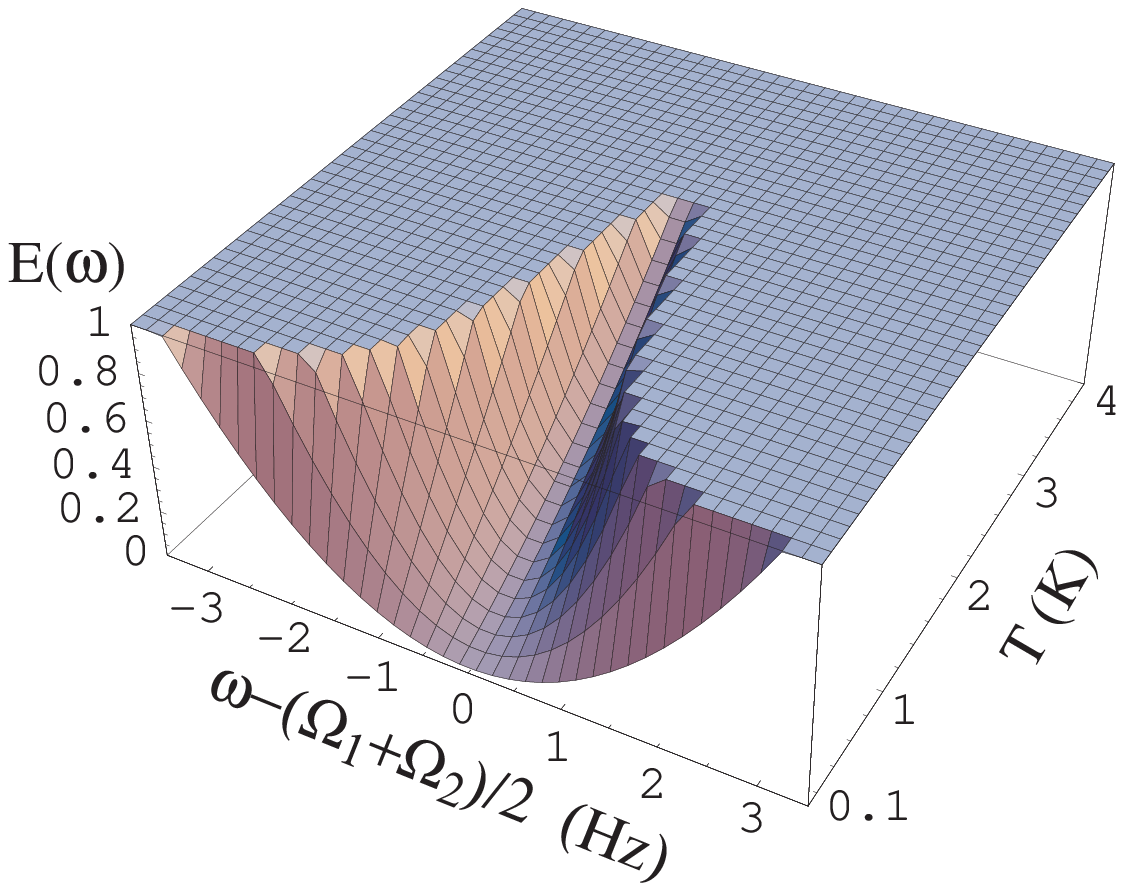,width=3.5in}}
\caption{ 
Degree of entanglement $E(\omega)$ 
of Eq.~(\protect\ref{degree}) versus 
frequency and temperature $T$, in the case of a mechanical frequency 
mismatch $\Omega_{2}-\Omega_{1}=10$ Hz.
The plot has been cut at $E(\omega)=1$.
The other parameter values are in the text.
}
\label{fig3}
\end{figure}

\begin{figure}[t]
\centerline{\epsfig{figure=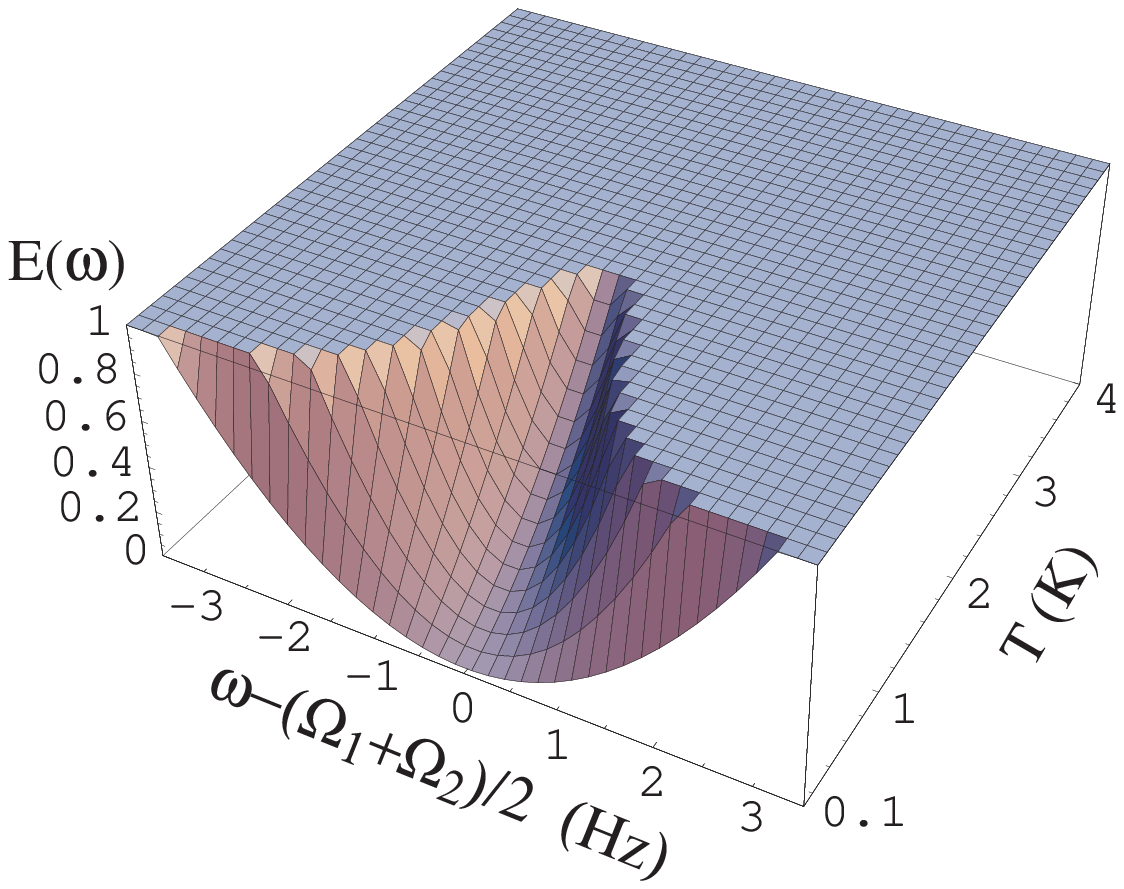,width=3.5in}}
\caption{
Degree of entanglement $E(\omega)$ 
of Eq.~(\protect\ref{degree}) versus 
frequency and temperature $T$, in the case of a mechanical frequency 
mismatch $\Omega_{2}-\Omega_{1}=20$ Hz.
The plot has been cut at $E(\omega)=1$.
The other parameter values are in the text.
}
\label{fig4}
\end{figure}

\end{document}